\documentclass[12pt]{report}

\usepackage{graphicx,amssymb,amsmath,setspace,multicol}
\usepackage{natbib209,sidecap,wrapfig,sectsty}
\usepackage[small,sc]{caption}
\usepackage[layout=modern]{advancedcoverpage}
\usepackage[usenames]{color}

\sectionfont{\normalsize}

  \renewenvironment{thebibliography}[1]{%
    \begin{oldthebibliography}{#1}%
      \footnotesize
      \setlength{\parskip}{0ex}%
      \setlength{\itemsep}{0ex}%
  }%
  {%
    \end{oldthebibliography}%
  }

\newcommand\aj{AJ}

\newcommand\apj{ApJ}
\newcommand\apjl{ApJ}

\newcommand\aap{A\&A}

\newcommand\mnras{MNRAS}

\newcommand\nat{Nature}

\newcommand\procspie{Proc.~SPIE}

\let\ga=\gtrsim
\newcommand\ion[2]{#1$\;${\small\rmfamily\@Roman{#2}}\relax}%
\let\ga=\gtrsim

\newcommand\afe{[$\alpha$/Fe]}

\newcommand\feh{[Fe/H]}
\newcommand\mathfeh{\mathrm{[Fe/H]}}

\sectionfont{\large}

\title{The Role of Dwarf Galaxies \\ in Building Large Stellar Halos}
\subtitle{A White Paper Submitted to the \\ Decadal Survey Committee}

\presentedto{\underline{Science Frontier Panels:} \\
{\it PRIMARY: The Galactic Neighborhood (GAN)} \\
{\it SECONDARY: Stars and Stellar Evolution (SSE)}}

\author{\emph{\underline{Authors}}\\
Evan~N.~Kirby \\
University of California Santa Cruz \\
(831) 459-3259 \\
\texttt{ekirby@ucolick.org} \\
~~ \\
Puragra~Guhathakurta, University of California Santa Cruz \\
James~S.~Bullock, University of California Irvine \\
Anna~Frebel, Harvard-Smithsonian Center for Astrophysics \\
Marla~Geha, Yale University \\
Karoline~M.~Gilbert, University of Washington \\
Jasonjot~S.~Kalirai, Space Telescope Science Institute \\
Manoj~Kaplinghat, University of California Irvine \\
Michael~Kuhlen, Institute for Advanced Study \\
Steven~R.~Majewski, University of Virgina \\
Brant~E.~Robertson, University of Chicago \\
Joshua~D.~Simon, Observatories of the Carnegie Institute of Washington \\
Marcel~Zemp, University of Michigan
}

\organization{\underline{Projects Emphasized:} \\
1. Thirty Meter Telescope, {\tt http://www.tmt.org} \\
2. Giant Magellan Telescope, {\tt http://www.gmto.org}} 
\date{}

\begin{document}
\maketitle



\noindent {\sc Abstract.}  The hierarchical theory of galaxy formation
rests on the idea that smaller galactic structures merge to form the
galaxies that we see today.  The past decade has provided remarkable
observational support for this scenario, driven in part by advances in
spectroscopic instrumentation.  Multi-object spectroscopy enabled the
discovery of kinematically cold substructures around the Milky Way and
M31 that are likely the debris of disrupting satellites.  Improvements
in high-resolution spectroscopy have produced key evidence that the
abundance patterns of the Milky Way halo and its dwarf satellites can
be explained by Galactic chemical evolution models based on
hierarchical assembly.

\vskip 0.3cm
\noindent
These breakthroughs have depended almost entirely on observations of
nearby stars in the Milky Way and luminous red giant stars in M31 and
Local Group dwarf satellites.  In the next decade, extremely large
telescopes will allow observations far down the luminosity function in
the known dwarf galaxies, and they will enable observations of
individual stars far out in the Galactic halo.  The chemical abundance
census now available for the Milky Way will become possible for our
nearest neighbor, M31.  Velocity dispersion measurements now available
in M31 will become possible for systems beyond the Local Group such as
Sculptor and M81 Group galaxies.  Detailed studies of a greater number
of individual stars in a greater number of spiral galaxies and their
satellites will test hierarchical assembly in new ways because
dynamical and chemical evolution models predict different outcomes for
halos of different masses in different environments.

\section{Introduction}
\label{sec:intro}

It is well established that hierarchical merging plays a key role in
galaxy formation.  Observational evidence ranges from dramatic major
mergers, such as the Antennae Galaxies, to subtler, low surface
brightness shells around galaxies like NGC~3923.  Evidence also
includes long tidal streams, such as the Sagittarius stream in the
Milky Way \citep{iba94}, the Giant Southern Stream in M31
\citep{iba01}, and large loops around NGC~5907 and other galaxies
\citep{mar-del08}.  \citet{bel08} found that the Milky Way (MW) halo
exhibits a large amount of substructure, and very recently, the degree
of kinematic substructure on the smallest scales has been quantified
in a systematic, statistical search of the inner MW halo (Schlaufman
et al., in preparation).

\citet{sea78} posited that the MW formed from the accretion and
dissolution of dwarf galaxies.  The dwarf spheroidal (dSph) galaxies
that exist today may be the lone survivors from the cannibalistic
construction of the Galactic halo.  However, chemical abundance
measurements in the last decade suggest that the stellar halo is not
chemically similar to many of its dSph satellites.  We address in
Section~\ref{sec:mismatch} how these dissimilarities affect the
paradigm of hierarchical assembly, which leads to a discussion of
modeling the chemical enrichment of dwarf galaxies in
Section~\ref{sec:models}.  In the last decade, groups in several
countries around the globe have focused much effort on spectroscopy of
stars in the MW, but studies in the coming decade will begin to probe
more distant galaxies.  The structure of M31 and its satellites
differs from the MW in a number of ways (addressed in
Section~\ref{sec:othergals}), which indicates that the MW may not be
representative of most galaxies.  In order to make robust comparisons
to theoretical predictions, a sample larger than one is necessary.  In
Section~\ref{sec:facilities}, we list some challenges for the next
decade's scientists.

\section{Comparing Observations of Stellar Abundances in Dwarf \\ Galaxies to the Milky Way Halo}
\label{sec:mismatch}

\begin{figure}[t!]
\centering
\includegraphics[width=0.5\textwidth]{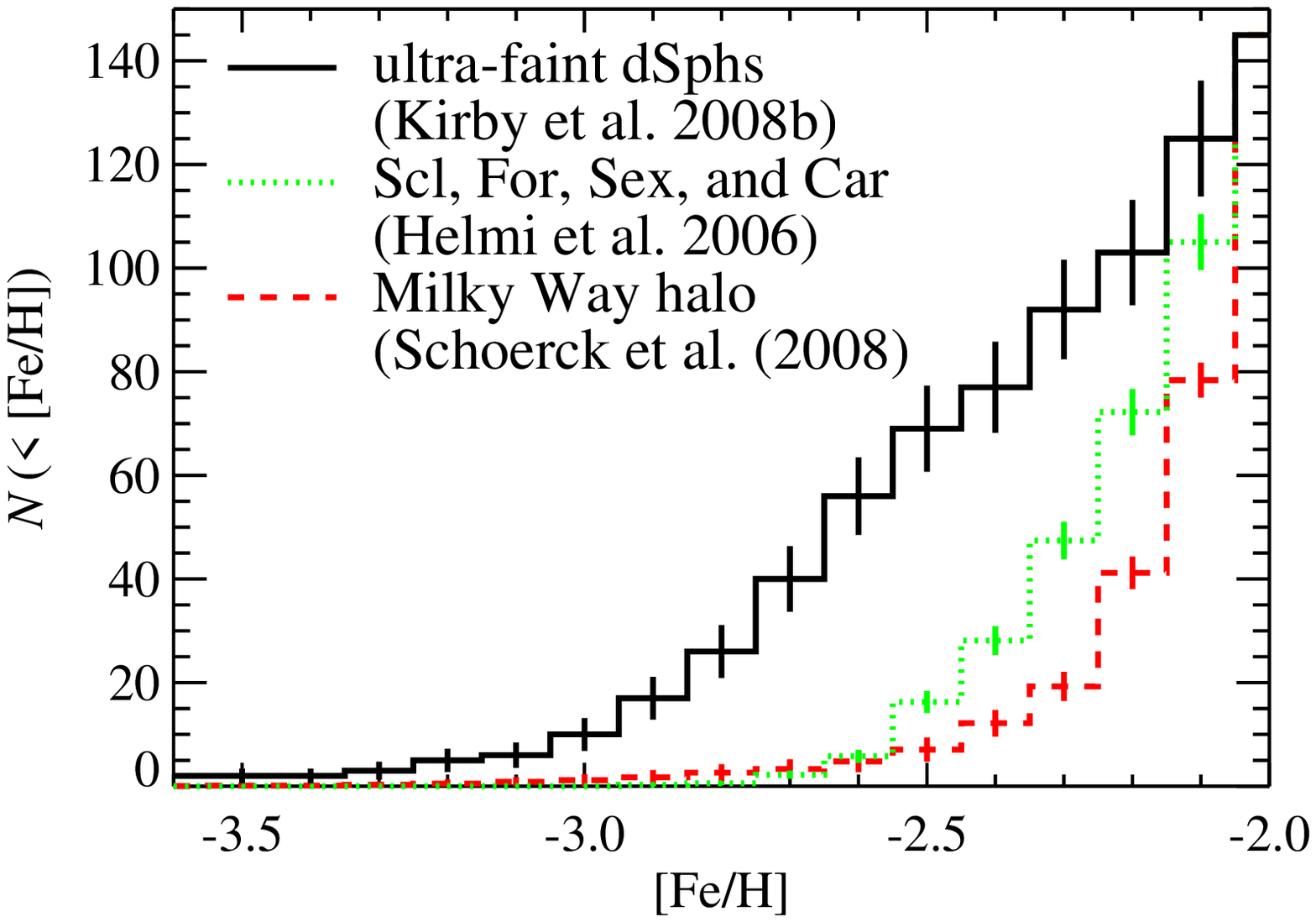}
\caption[]{Observed cumulative metallicity distributions for dSphs and
  the MW halo, normalized to match at $\mathfeh = -2$.  The red line
  represents the halo measurements from the Hamburg/ESO Survey.  The
  green line shows measurements for fairly luminous MW satellites.
  The black line shows \feh\ for eight of the faintest MW dSphs.  The
  halo seems to have metal-poor counterparts in dSphs at all
  metallicities, but the shape of the dwarf galaxy MDFs is
  significantly weighted toward low \feh\ compared to the
  halo.\label{fig:mdf}}
\end{figure}

{\bf In the past decade, we learned that stars in dwarf galaxies have
compositions different from field stars in the MW halo.}
\citet{hel06} measured the metallicity distribution functions (MDFs)
of some of the more luminous dSphs.  The metal-poor tails of the MDFs
of the most luminous dSphs (e.g., Fornax) appear similar to that of
the MW halo \citep{sch08}.  However, the MDFs of less luminous dSphs
are more metal-poor.  In fact, \citet{kir08b} found much more
metal-poor MDFs in \citeauthor{sim07}'s (\citeyear{sim07})
spectroscopic sample of eight of the least luminous MW satellites,
including a significant population of extremely metal-poor (EMP,
$\mathfeh < -3$) stars.  Figure~\ref{fig:mdf} summarizes the results
from these three studies.

\begin{figure}[t!]
\begin{minipage}{0.4235\textwidth}
\centering
\includegraphics[width=0.95\textwidth]{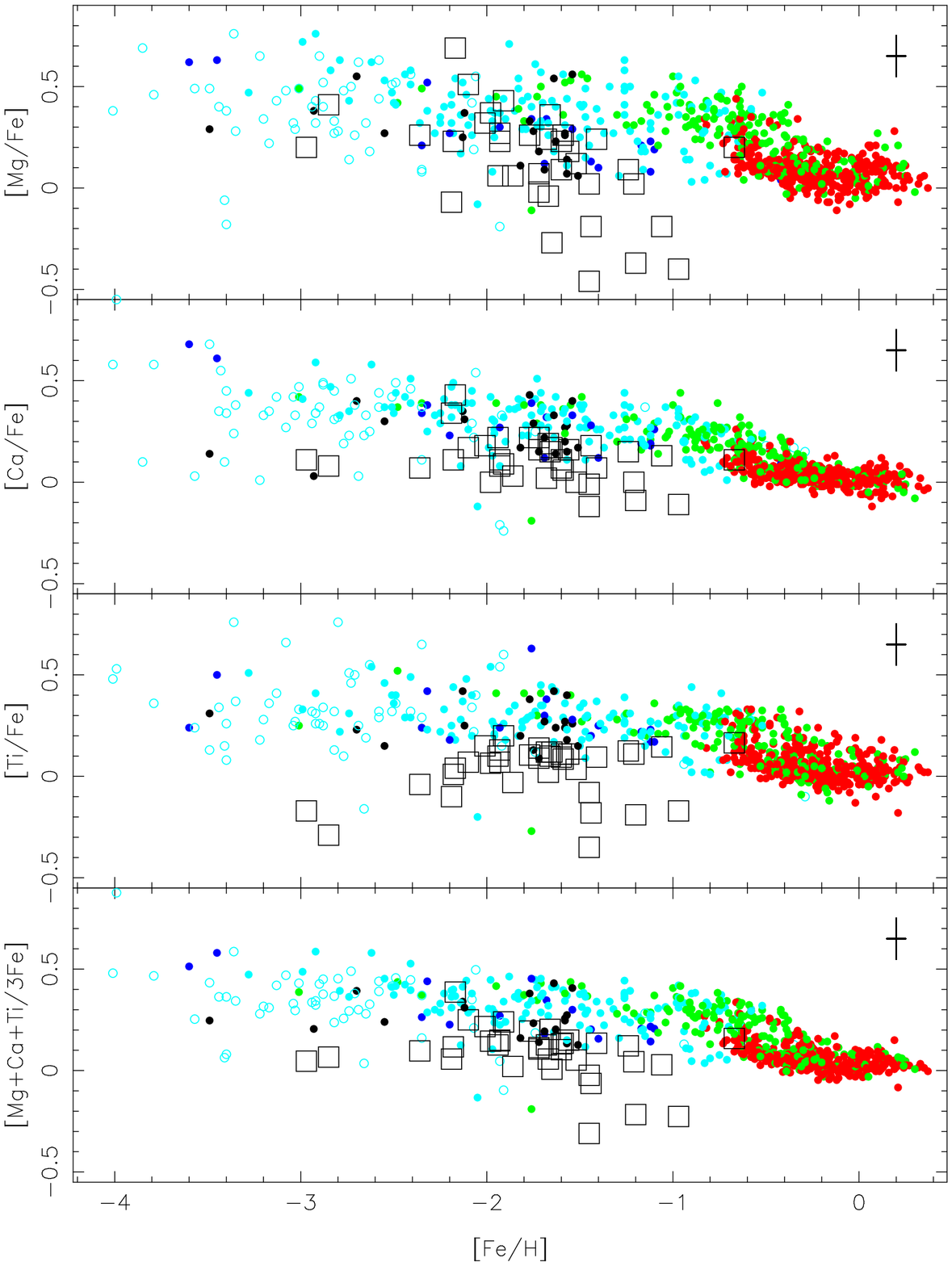}
\caption[]{Compilation of abundances \citep{ven04} in dSphs
  ({\it black squares}) and different MW components: thin disk ({\it
  red}), thick disk ({\it green}), halo ({\it cyan}), high velocity
  ({\it black}), and retrograde ({\it blue}).  \afe\ is lower at a
  given \feh\ in dwarf galaxies than in the halo.\label{fig:ven04}}
\end{minipage}
\quad
\begin{minipage}{0.5465\textwidth}
\centering
\includegraphics[width=\textwidth]{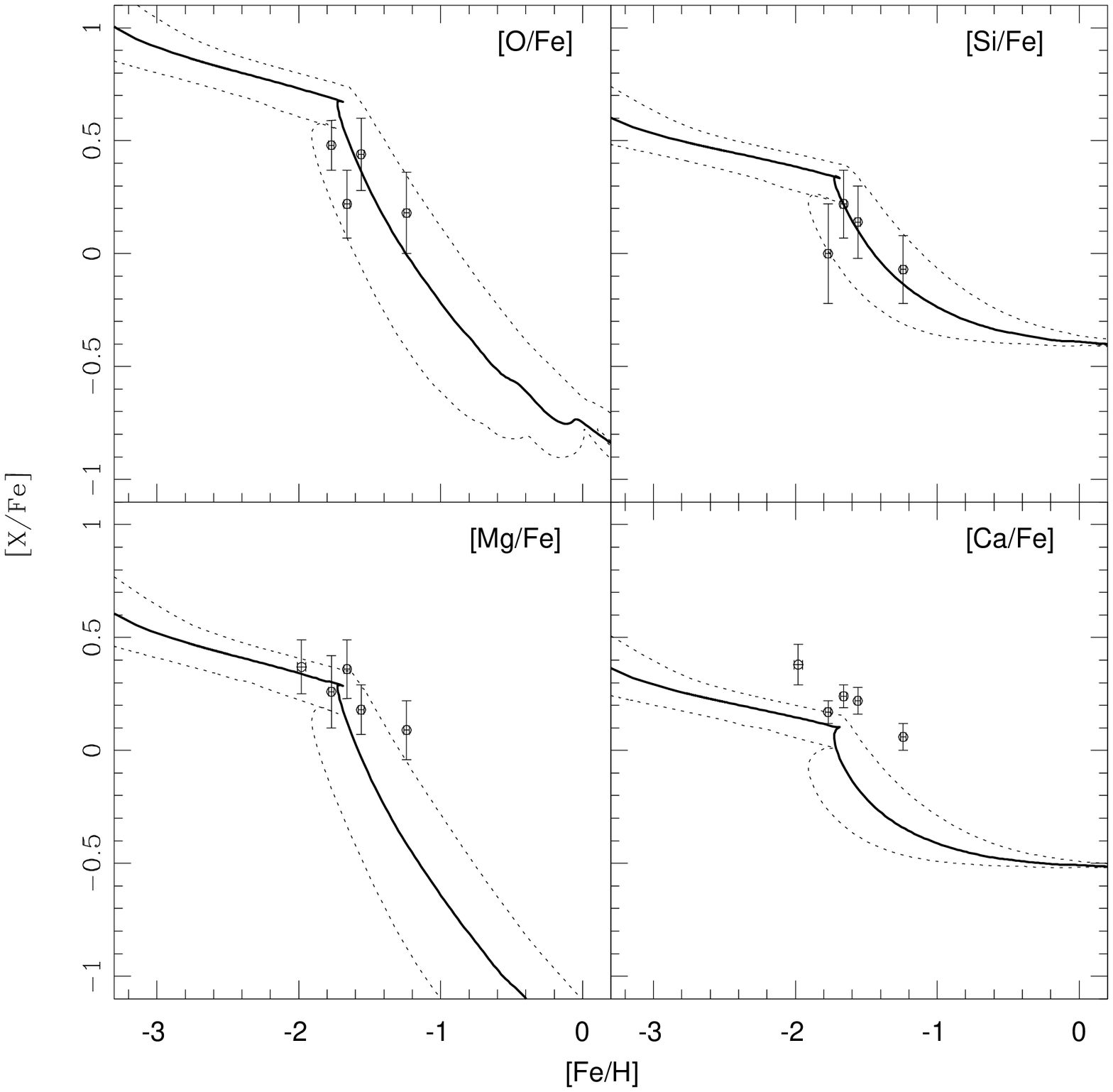}
\caption[]{The predictions of $\alpha$ element abundance ratios vs.\
  \feh\ for the Sculptor dSph \citep{lan04}.  O and Mg are produced
  primarily in Type~II supernovae whereas Si and Mg are produced in
  both Type~Ia and II supernovae.  As gradual star formation proceeds
  in a stellar system, \feh\ increases and \afe\
  decreases.\label{fig:lan04}}
\end{minipage}
\end{figure}

The discrepancies between the halo and dSphs also extend to elements
other than Fe.  The abundances of $\alpha$ elements, such as O, Mg,
Si, Ca, and Ti, together with Fe can determine the duration of star
formation in a stellar system.  In the beginning of the past decade,
high resolution spectroscopy of individual stars in dSphs showed that
the ratio \afe\ in dSphs is lower at a given \feh\ than in the halo
\citep[][see Fig.~\ref{fig:ven04}]{she01,she03,gei05}.

\vskip 0.3cm
\noindent
{\bf In the next decade, we will discover new patterns in known and
soon-to-be-discovered dwarf galaxies.}  The ultra-faint dwarf galaxies
appear to harbor relatively large populations of the lowest
metallicity stars.  Detailed chemical abundances in these systems
\citep{fre09} provide observational constraints on early star
formation and the possibility that dwarf galaxies are the building
blocks of the MW.  This work is challenging because the targets are
very faint for high-resolution spectroscopy ($V \ga 17$).  Current
8--10~m telescopes can obtain spectra with adequate signal-to-noise to
measure detailed chemical abundances of the brightest stars in these
galaxies.  Existing surveys like \textcolor{Red}{\bf SEGUE}
\citep{yan09} and future photometric surveys like
\textcolor{Red}{\bf LSST} \citep{tys02}, \textcolor{Red}{\bf
SkyMapper} \citep{mur08}, and \textcolor{Red}{\bf Pan-STARRS}
\citep{kai02} will discover more tiny galaxies.  Reaching more stars
and systems farther away will require a new generation of large
aperture, ground-based telescopes such as \textcolor{Red}{\bf
TMT} and/or \textcolor{Red}{\bf GMT}.

Multi-dimensional abundance measurements \citep[using the technique
of, e.g.,][]{kir08a} for large samples of stars have very recently
become available via medium-resolution spectroscopy on the largest
telescopes (27 stars in the Leo~II dSph, \citeauthor{she09}\
\citeyear{she09}, and $\sim 400$ stars in the Sculptor dSph,
\citeauthor{kir09} \citeyear{kir09}).  In the next few years, this
technique, based on spectral synthesis, will permit \afe\ and \feh\ to
be measured simultaneously for 100--1000 red giants in many of the
brighter MW dSphs.  However, in the most distant dSphs (e.g., Leo~I
and Leo~T), spectroscopy with current 8--10~m telescopes is practical
for individual stars only on the upper part of the red giant branch
(RGB).  Further down the RGB or in more distant stellar systems, the
medium-resolution spectra with 8--10~m telescopes must be coadded to
attain the requisite signal-to-noise.

\section{Modeling Chemical Enrichment in Dwarf Galaxies}
\label{sec:models}

{\bf In the last ten years, we learned that chemical discrepancies
  between dwarf galaxy stars and halo field stars are natural
  consequences of $\Lambda$CDM cosmology.}  At first glance, these
discrepancies seem to challenge the assertion that the stellar halo
formed mostly or entirely from dwarf galaxies.  In fact, all of these
discrepancies can be explained by cosmologically motivated models of
star formation and chemical enrichment.

The observed chemical abundance patterns in dSphs can be explained by
self-contained star formation models \citep{lan04,mar06,mar08}.  As
star formation proceeds, \feh\ increases, but after $\sim 1$~Gyr,
Type~Ia supernovae begin to lower \afe.  Therefore, the ``knee'' in
the \afe--\feh\ diagram indicates the metallicity reached by about
1~Gyr, which in turn indicates the vigor of star formation.
Figure~\ref{fig:lan04} shows \citeauthor{lan04}'s (\citeyear{lan04})
model of the Sculptor dSph.

In these models, galactic winds regulate the rate of star formation.
Less massive dSphs are more susceptible to losing gas to winds and
therefore have slower, less efficient star formation.  Because rapid
star formation increases \feh\ quickly and keeps \afe\ elevated,
smaller dSphs tend to have lower mean \feh\ than larger dSphs, and an
\afe\ that begins to decline at lower metallicity.  Therefore,
Fig.~\ref{fig:ven04} immediately indicates that the dSphs that built
the majority of the MW halo were more massive than the surviving
dSphs.

\begin{figure}[t!]
\centering
\includegraphics[width=\textwidth]{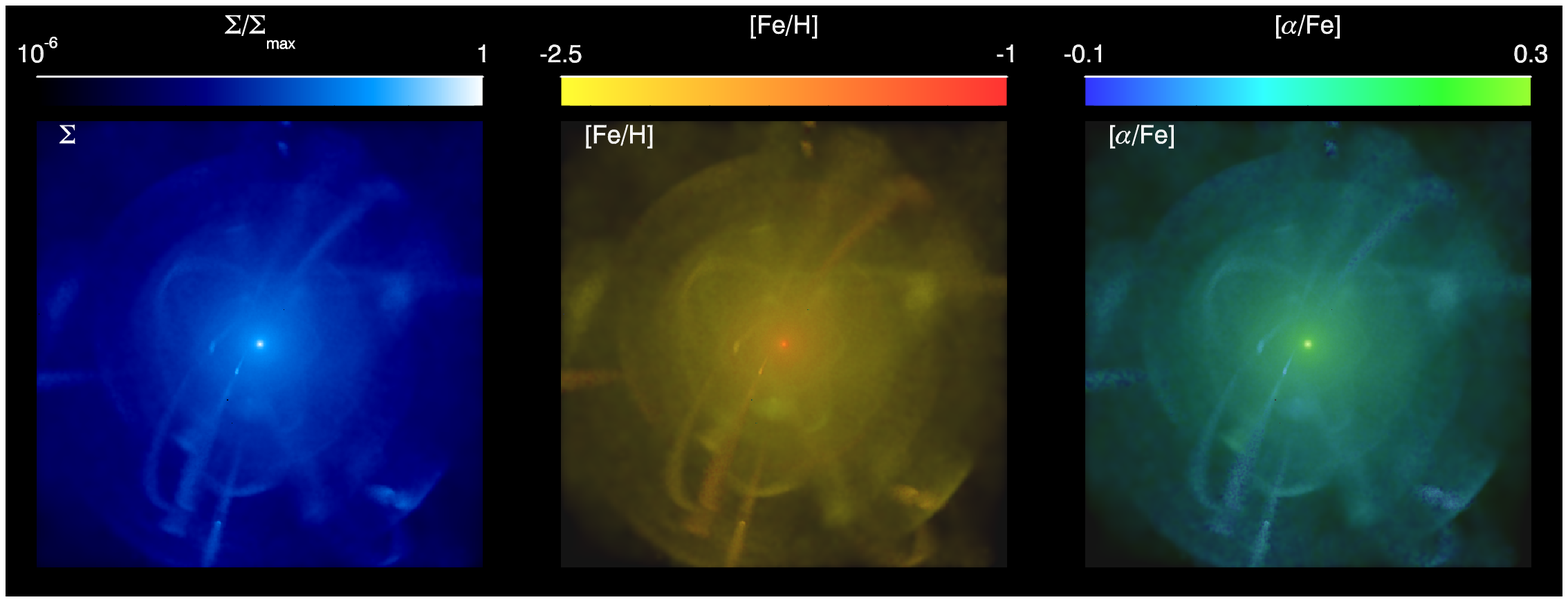}
\caption[]{Simulation \citep{fon06a} of a Milky Way-sized system
  showing (left to right) surface brightness, metallicity, and \afe.
  The box is 200~kpc on a side.  In this model, the early, rapid
  construction of the diffuse stellar halo from relatively massive
  dwarf galaxy progenitors imprints higher \feh\ and \afe\ abundances
  than in the cold substructure.\label{fig:halo}}
\end{figure}

As \citet{rob05} pointed out, this observation is completely
consistent with cosmologically motivated models of galaxy formation
\citep[e.g.,][]{bul05}.  In dark matter simulations, a few relatively
massive satellites built the bulk of the halo $\sim 10$~Gyr ago.
Surviving dSphs continue to add stars to the halo via tidal
disruption, but those dSphs are not representative of a typical halo
star.  Figure~\ref{fig:halo} shows the zero-redshift view of one of
these MW-sized simulations with a prescription for star formation.  In
the context of a hierarchical formation scenario for the MW stellar
halo and substructure, the surviving cold components of the halo,
including dSphs and debris streams, may exhibit lower \feh\ and lower
\afe\ than the inner stellar halo owing to their low masses and
extended star formation histories.  The inner halo may have been
constructed from more massive, rapidly-forming systems that have
comparably enhanced \feh\ and \afe\ abundances.

\vskip 0.3cm
\noindent
{\bf In the coming decade, synergy between model predictions and
observations will advance our understanding of the origin of the MW
halo.}  It is difficult to directly observe the progenitors of the
bulk of the halo because they have already been destroyed.  However,
it is possible to connect the properties of surviving dSphs and
stellar halos using models of galactic star formation and chemical
enrichment.  These simulations make predictions about the kinematics
\citep{hel99,fon06b} and chemistry \citep{fon06a} of the relics of
accretion events.  Confirmation of these predictions will require
coordinated spectroscopic surveys focused on measuring radial
velocities and multi-dimensional abundances for a large number of
stars.  One such survey targeting the Galactic bulge is
\textcolor{Red}{\bf APOGEE} \citep{all08}.

The number of stellar chemical abundance measurements in dwarf
galaxies and the MW halo is steadily increasing, and extremely large,
ground-based telescopes will reach further down the luminosity
function, rapidly multiplying that number.  The quality of chemical
evolution models will evolve with the quality of the data.  In
particular, we require a working model of star formation in the
tiniest galaxies.  Recent evidence for a threshold galaxy mass for
star formation \citep{str08} begs for a theoretical explanation that
could also account for the abundance patterns in tiny dwarf galaxies.
Can the same model also explain the abundances of EMP halo field
stars, supporting the hypothesis that they originated in tiny dwarf
galaxies?  These observations will be complemented by star formation
studies from space-based photometry (e.g., \textcolor{Red}{\bf
JWST}).

As more data becomes available, abundance patterns may be scrutinized
in finer detail.  Forthcoming high-resolution spectroscopy will
increase the body of data on neutron-capture elements in dwarf
galaxies and the halo.  Given the discrepancies in $\alpha$ elements
between the dwarf galaxies and the halo, it would not be surprising to
find that the abundances of heavier elements also diverge.  A chemical
evolution model of those elements will be required to explain these
observations.  Can the same chemical evolution model describe the
heavy element patterns in both dwarf galaxies and the halo?  Or will
the heavy element abundances teach us that at least some halo stars
could not have been born in dwarf galaxies?

\section{Galaxies and Environments Are Not All the Same}
\label{sec:othergals}

{\bf Recent observations of M31 have taught us that the MW system is
not representative of all spiral galaxies.} At first sight, the MW and
M31 are both examples of typical late-type galaxies, but closer
inspection elucidates subtle but important differences.  For example,
M31 seems to have had a more violent merger history than the MW
\citep{iba01,guh05}.  Furthermore, the demographics of the satellite
populations are different.  M31, unlike the MW, possesses no known
dwarf irregular galaxies.  The MW, unlike M31, possesses no dwarf
elliptical galaxies.  Even the surviving M31 dSphs have different
structural parameters than their MW counterparts \citep{mcc06,pen08}.

Another powerful prediction of the simulations discussed in
Section~\ref{sec:models} is the difference between the MW and galaxies
of different sizes.  \citet{fon08} and \citet{gil09} test the
metallicity predictions of these simulations on observations of M31,
and the same models could make predictions about how \afe\ in the M31
system differs from the MW system.  M33 and NGC~55 offer even more
dramatic comparisons.  (NGC~55 has the advantage of being face-on for
easier disk-spheroid separation.)  Although both are star-forming,
spiral galaxies, their masses are much smaller than and have different
gas content from the MW or M31.  Cosmological simulations would
predict measurably distinct kinematic and abundance patterns for the
halos and dwarf galaxies of M33 and NGC~55 compared to the MW or M31.
Contrasting predictions can be made even between M33 and NGC~55 based
on their broader environments.  M33 lies close to M31 whereas NGC~55
is a likely companion of NGC~300, a much smaller galaxy than M31---and
also a good candidate for future medium-resolution spectroscopy.  The
many galaxies in the Sculptor and M81 Groups would provide even more
data for this comparison if they were observable.  These groups
exhibit different galaxy mass distributions, velocity dispersions, and
intra-group gas densities, all of which can yield different, testable
predictions in numerical simulations of galaxy assembly.

\vskip 0.3cm
\noindent
{\bf In the next decade, we will test in detail hierarchical assembly
models in galaxies beyond the MW.}  Many predictions for M31 and the
Sculptor and M81 Groups are not testable with present observational
capabilities.  Kinematic measurements of individual stars are limited
to the MW and M31, and abundance measurements of individual stars are
limited to the MW and its satellites.  Extremely large telescopes of
the next decade may greatly increase the distance (and hence the
variety of galaxies) to which these measurements are possible.
Spectra of individual stars in M31 with resolving power of $R \sim
6000$ and signal-to-noise of $\sim 30~{\mathrm \AA}^{-1}$ are feasible
with the next generation of optical telescopes.  With longer
exposures, even the tip of the RGB in NGC~55 would be reachable with a
spectral quality capable of yielding multi-dimensional abundances.
With spectral coaddition of individual stars, galaxies in the Sculptor
and M81 Groups would become accessible.  These spectroscopic
observations will rely on wide-field photometric surveys, such as
\textcolor{Red}{\bf LSST}, \textcolor{Red}{\bf Pan-STARRS},
and \textcolor{Red}{\bf SkyMapper}, which will accurately
characterize the level of substructure and discover new satellites
around galaxies other than the MW.  Ground-based spectroscopy will
also complement observations from space-based telescopes, such as
\textcolor{Red}{\bf JWST}, which will measure the star formation
histories of distinct components of galaxies more distant than the MW
and M31 \textcolor{Red}{(see the white paper by T.~Brown et al.,
``The History of Star Formation in Galaxies'')}.


\section{Summary Goals for the Next Decade}
\label{sec:facilities}

In order to test $\Lambda$CDM formation scenarios thoroughly, their
predictions must be verified in a wide range of large stellar halos
and dwarf galaxies in various stages of disruption.  Cosmologically
motivated simulations and star formation models make different
predictions about the kinematic and chemical profiles of dSphs based
on their sizes and distances; of tidal streams based on their surface
brightnesses; and of diffuse stellar halos based on the total galaxy
mass.  Presently, the only components with observationally accessible
kinematics and chemistry are nearby dSphs; high-surface brightness or
very nearby tidal streams; and the MW and M31 halos.  The future of
these observations will depend on both new facilities and rethinking
the ways in which we use existing facilities.

{\bf Existing facilities:} In order to expand the body of
spectroscopic data in dwarf galaxies using current 8--10~m telescopes,
several international groups have turned to medium-resolution
spectroscopy.  Recent refinements to abundance measurement techniques
\citep{she09,kir09} make the most of this data.  Mining data for new
discoveries is, in fact, the approach that led to the recent
discoveries of the tiniest known galaxies \citep[e.g.,][]{bel07}.  New
ways of exploiting existing data and existing facilities should not be
underestimated as a major mode for new discoveries in the next ten
years.

{\bf New facilities:} An extremely large, ground-based telescope, such
as \textcolor{Red}{\bf TMT} and/or \textcolor{Red}{\bf GMT},
would increase the body of kinematic data for galactic systems.  It is
important to measure the velocity dispersions and masses of dwarf
galaxies in order to compare them to their simulated representatives.
The largest telescopes today are barely able to measure velocity
dispersions of the smallest dSphs around M31, and those measurements
are sometimes based on fewer than 10 stars.  Because only a handful of
stars in each dSph are bright enough for spectroscopy, radial velocity
surveys have almost certainly missed many dSphs around both the MW and
M31.  Furthermore, no dSph beyond the MW and M31 systems is near
enough to permit measurements of stellar velocity dispersion.

\textcolor{Red}{\bf TMT} and/or \textcolor{Red}{\bf GMT}
could reach more, fainter stars in known dSphs and probe larger
distances to discover new ones.  The next decade will witness a leap
in the amount of high- and medium-resolution spectroscopic data that
can be used to support or disprove the subtleties of $\Lambda$CDM
galaxy formation.  Measurements within the MW and M31 will become more
refined, and galaxies beyond the Local Group will become laboratories
for studying the dynamical and chemical processes that transform small
galaxies into large ones.  The census of dSphs around the MW will
become more complete, and multi-dimensional abundance studies will
shed light on the formation of M31's stellar halo and the stellar
halos of galaxies in the Sculptor and M81 Groups.

\begin{multicols}{2}
\renewcommand\bibsection{}

\end{multicols}

\end{document}